\documentclass[twocolumn,twoside,slac_two]{revtex4}
\usepackage{graphicx}
\usepackage{fancyhdr}
\usepackage{relsize}

\def\mmbc{M^{}_{\rm bc}}

\def\ra{\!\rightarrow\!}
\def\bbar{\overline{B}{}^{\,0}}

\def\arhorho{${\cal A}$}
\def\srhorho{${\cal S}$}

\def\babar{\mbox{\slshape B\kern-0.1em{\smaller A}\kern-0.1em
    B\kern-0.1em{\smaller A\kern-0.2em R}}}

\begin{document}

\title{Measurement of the CKM angle {\boldmath  $\phi_2 (\alpha)$}\footnote{University 
of Cincinnati preprint number UCHEP-07-07.}}

%

\author{A. Somov}
\affiliation{University of Cincinnati, Cincinnati, Ohio 45221, USA}

\begin{abstract}
We present recent measurements of the unitarity triangle angle 
$\phi_2$ ($\alpha$) using $B\rightarrow \pi\pi$, $B\rightarrow \rho\rho$, 
and $B\rightarrow \rho\pi$ decays. The measurements are based on data 
samples collected with the Belle and BaBar detectors at the KEKB and PEP-II 
$e^+ e^-$ colliders, respectively. We also report on a new measurement of a $CP$-violating
asymmetry in $B^0 \rightarrow a_1^\pm\pi^\mp$ decay which will allow 
to constrain further the angle $\phi_2$.
\end{abstract}

\maketitle

\thispagestyle{fancy}


\section{Introduction}
\begin{figure}[h]
\centering
\includegraphics[width=25mm,angle=-90]{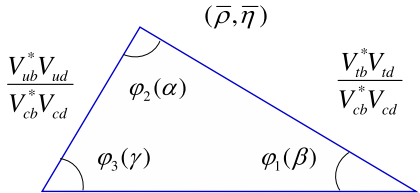}
\caption{The CKM unitarity triangle
$V_{ud}V^*_{ub} + V_{cd}V^*_{cb} + V_{td}V^*_{tb} = 0$ in the complex
$\overline{\rho}-\overline{\eta}$ plane.}
\label{fig:triangle}
\end{figure}

\begin{figure}[h]
\centering
\includegraphics[width=40mm]{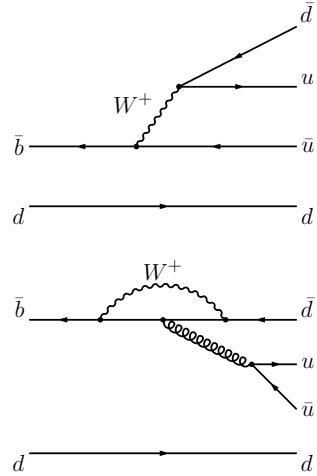}
\caption{Tree-level (top) and gluonic penguin (bottom) diagrams for the
decays $B^0\rightarrow \pi^+\pi^-,\;\rho^+\rho^-,\;\rho^\pm\pi^\mp,\;a_1^\pm\pi^\mp$.}
\label{fig:feynman}
\end{figure}

The $CP$ violation in the standard model (SM) can be described
by the presence of a complex phase in the three-generation 
Cabibbo-Kobayashi-Maskawa~\cite{ckm} (CKM) quark-mixing matrix. Unitarity 
constraints on the matrix elements lead to six relations, one of
the most interesting is $V_{ud}V^*_{ub} + V_{cd}V^*_{cb} + V_{td}V^*_{tb} = 0$.
This relation can be depicted as a triangle in the complex plane as shown in
Fig~\ref{fig:triangle}. Checking unitarity of the CKM 
matrix implies measuring sides and angles of the triangle. This provides 
an important test of the SM. The phase angle $\phi_2$~\cite{angle_definition}, defined 
as ${\rm arg}[-(V_{td}V_{tb}^*)/(V_{ud}V_{ub}^*)]$, represents the phase 
difference between $V_{td}$ and  -$V^*_{ub}$. It can be determined by measuring a 
time-dependent $CP$ asymmetry in charmless $b\rightarrow u\overline{u}d$ decays 
such as $B^0\rightarrow\pi^+\pi^-,\,\pi^+\pi^-\pi^0,\,\rho^+\rho^-$, and 
$a_1^\pm\rho^\mp$  ~\cite{chargeconjugate}. The decay-rate asymmetry in these 
decays can arise due to the interference between the amplitudes of the direct 
decay of $B$ and decay after $B\overline{B}$  mixing. The $B$ decays are proceeded 
mainly through a tree and gluonic penguin loop diagrams as shown in Fig~\ref{fig:feynman}.
The penguin loop amplitude is irrelevant to the $\phi_2$ and contaminates
the measurement. The penguin contribution can be constrained by using isospin 
relations or employing SU(3) flavor relations, which will be discussed later.

The first  analysis of time-dependent $CP$-violating asymmetries 
has been performed using $B^0\rightarrow \pi^+\pi^-$ decays. This relatively 
clean channel provides hight-precision measurements but has a significant 
contribution from a penguin loop amplitude. This is indicated by observation of a 
large direct $CP$ violation by Belle and by measurement of a relatively large 
branching fraction of $B^0\rightarrow \pi^0\pi^0$ decay. The measurements of 
$CP$ asymmetries in $B^0\rightarrow \pi^+\pi^-$ are followed by that in 
$B^0\rightarrow \rho^+\rho^-$ decays. The contamination from a $b\rightarrow u$ 
penguin amplitude is measured to be much smaller here. However, a large width of 
$\rho$ mesons makes reconstruction of $B^0\rightarrow \rho^+\rho^-$ decays a challenging task.  
The extraction of $\phi_2$ from measurements in $B^0 \rightarrow \pi^+\pi^-$ and 
$B^0 \rightarrow \rho^+\rho^-$ decays can be performed using an isospin analysis which
allows one to constrain the  contribution from the penguin amplitude generally with an 
eight-fold ambiguity. The latest measurements in $B^0\rightarrow \pi^+\pi^-$ and 
$B^0\rightarrow \rho^+\rho^-$ decays are discussed in Section~\ref{sec:pipi} and 
\ref{sec:rhorho}.

Snyder and Quinn~\cite{snyder_quinn} showed that the angle 
$\phi_2$ can be determined without discrete ambiguities using a time-dependent Dalitz 
plot (DP) analysis of $B^0\rightarrow \pi^+\pi^-\pi^0$ decays. The DP analysis allows 
one to measure the complex amplitudes of  $\pi^+\pi^-\pi^0$ decays which are related 
to $\phi_2$ via an isospin relation. Previously a quasi-two-body analysis of 
$B^0\rightarrow\rho^+\pi^-$ decays has been used to constrain $\phi_2$ using this 
decay mode. Recent results on the DP analyses are presented in Section~\ref{sec:rhopi}.

$B^0\rightarrow a_1^\pm\pi^\mp$ is another channel which allows us to 
extract the angle  $\phi_2$. Recently BaBar has measured a  $CP$-violating asymmetry in 
this decays. However the extraction of $\phi_2$ has not yet been performed 
for this channel. It is expected to be done using a flavor SU(3) symmetry. 
The measurements in $B^0\rightarrow a_1^\pm\pi^\mp$ decays are described 
in Section~\ref{sec:a1pi}.

\section{Analyses overview}
Analyses of $B\rightarrow \pi\pi$, $B\rightarrow \rho\pi$, $B\rightarrow \rho\rho$, 
and $B\rightarrow a_1\pi$ decays have several common features.

Charged pions are identified using information from time-of-flight counters and 
the central tracking chambers. Most analyses also use information from aerogel 
threshold cherenkov counters (Belle) and from a DIRC detector (BaBar).
$B$ decays are selected by combining  charged pions originating from 
the interaction region, and adding $\pi^0$'s  in the decays which 
involve $\rho^\pm$ mesons.

$B$ decays are identified using two kinematic variables:
the beam-energy-constrained mass 
$\mmbc\!\equiv\!\sqrt{E^2_{\rm beam}-p^2_B}$ and energy difference
$\Delta E\!\equiv\!E^{}_B-E^{}_{\rm beam}$, where $E^{}_{\rm beam}$ is the
beam energy, and $E^{}_B$ and $p_B$ are the energy and momentum of the
reconstructed $B$  candidate, all evaluated in the center-of-mass (CM) frame.

Flavor of the $B$ meson accompanying the signal $B$ is identified using 
a tagging algorithm which identifies  its decay products (mainly leptons 
and kaons). The  tagging algorithm provides the flavor of the tagged meson 
and a tagging quality.

The dominant background originates from  $e^+e^-\!\rightarrow q\bar{q}\ (q=u,d,s,c)$
continuum events. The separation of continuum background from the signal events 
can be done using the event topology: $q\bar{q}$ events tend to be more 
jet-like while  $B\overline{B}$ events are more spherical in the CM frame.

The Belle analyses use event-shape variables, specifically, 16 modified
Fox-Wolfram moments~\cite{fox_wolfram} combined into a Fisher discriminant~\cite
{KSFW}. We form signal and background likelihood functions ${\cal L}^{}_s$ and
${\cal L}^{}_{BG}$ by multiplying a probability density function (PDF) for the 
Fisher discriminant by a PDF for $\cos\theta^{}_B$, where $\theta^{}_B$ is the polar 
angle in the CM frame between the $B$ direction and the beam axis. The PDFs for signal 
and $q\bar{q}$ are obtained from Monte Carlo (MC) simulations and the data sideband, 
respectively.  We calculate the ratio ${\cal R} ={\cal L}^{}_s/({\cal L}^{}_s + {\cal L}^{}_{BG})$.
The continuum background is reduced by setting a threshold on $\mathcal R$. In 
$B^0\rightarrow \rho^+\rho^-$ analysis PDFs for $\mathcal R$ are included into 
a likelihood function. In some analyses we also set a threshold on the absolute value 
of the cosine of the angle between a thrust axis of the reconstructed signal $B$ meson
and that of the rest of the event, ${\rm cos_{th}}$. 
 
In the BaBar analyses the $q\bar{q}$ background is usually removed by directly 
applying thresholds on ${\rm cos_{th}}$ and on the second-to-zeroth Fox-Wolfram 
momentum $R_2$. Further discrimination is achieved by using an artificial neural 
network which includes event shape variables. The neural net is trained using MC 
simulated events and off-peak data. The output of the net, ${\mathcal N}$ is 
included as a PDF into the fitting likelihoods.

The decay time difference $\Delta t$ between the two $B$ mesons can be determined by
measuring the distance between the decay vertices of these mesons. Since the $B^0$ 
and $\bbar$  are produced approximately at rest in the $\Upsilon(4S)$ CM system, 
$\Delta t \simeq \Delta z / \beta \gamma c$, where $c$ is the speed of light and 
$\beta \gamma$ is a  Lorentz boost of the $\Upsilon(4S)$ mesons and equals 0.425 (0.559)
for Belle (BaBar). The decay vertices of the signal and tag-side $B$ mesons are 
reconstructed by fitting charged tracks that have hits in the silicon vertex detector
using an interaction point constraint.

\section{$B^0\rightarrow \pi^+\pi^-$ \label{sec:pipi}}
The time-dependent rate for $B \rightarrow \pi^+\pi^-$  decays tagged with $B^0$($Q=+1$) 
and $\bbar$ ($Q = -1$) mesons is given by
\begin{eqnarray}
\mathcal{P}_{\pi\pi} (\Delta t) =  \frac{e^{-|\Delta t|/\tau_{B^0}}}{4\tau_{B^0}}
\{ 1  + Q[{\cal A}_{}\cos(\Delta m \Delta t) \\
 + {\cal S}_{}\sin(\Delta m \Delta t)]\},\nonumber
\label{eqn:rate}
\end{eqnarray}
where $\tau_{B^0}$ is the $B^0$ lifetime, $\Delta m$ is the mass difference between
the two $B^0$ mass eigenstates, $\Delta t$ is the proper-time difference between
the two $B$ decays in the event, and ${\cal A_{}}$~\cite{asym_def}  and ${\cal S_{}}$ 
are $CP$ asymmetry coefficients which are to be obtained  from a fit to the data. 
If the decay amplitude is  dominated by a tree diagram, 
${\cal S}_{} = {\rm sin}(2\phi_2)$ and ${\cal A}_{} = 0$. The presence 
of an amplitude with a different weak phase (such as from a gluonic penguin diagram) gives rise 
to direct $CP$ violation and shifts ${\cal S}_{}$ from ${\rm sin}(2\phi_2)$: 
\begin{eqnarray}
{\cal S}_{\rm meas} = \sqrt{1 - {\cal A}_{\rm meas}^2} {\rm sin}2\phi_2^{\rm eff},
\label{eq:phi2eff}
\end{eqnarray} 
where ${\cal A,S}_{\rm meas}$ are measured coefficients, 
$\phi_2^{\rm eff} = \phi_2 + \delta \phi_2$, and $\delta\phi_2$ is the phase shift. 

\begin{figure}[h]
\centering
\includegraphics[width=35mm,angle=-90]{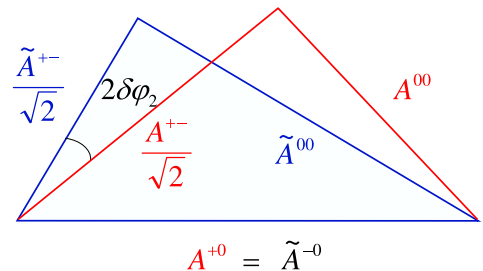}
\caption{The isospin triangles for $B\ra\pi\pi$ decays. The amplitudes $\widetilde{A}$ are
defined as $\widetilde{A}=e^{i2\phi_3}\bar{A}$. }
\label{isospin_triangle}
\end{figure}

The Belle analysis~\cite{pipi_belle} is based on a data sample consisting of 535 million 
$B\overline{B}$ pairs.
The analysis is organized in two steps. We first determine the yields of signal and 
background components using an unbinned extended maximum likelihood (ML) fit to 
$\mmbc$, $\Delta E$, and the kaon identification probability $x_{\pm}$ for the positively
and negatively charged tracks. The fit yields $1464\pm 65$ $\pi^+\pi^-$ candidates.
We subsequently perform a fit to the $\Delta t$ distribution for the $CP$ parameters  
${\cal A_{}}$ and ${\cal S_{}}$. The fit to 16831 events yields 
${\cal A_{\pi^+\pi^-}} = 0.55\pm0.08(stat)\pm 0.05(syst)$ and 
${\cal S_{\pi^+\pi^-}} = -0.61\pm0.10(stat)\pm 0.04(syst)$.

The BaBar measured the $CP$-violating parameters using a sample of 383 million 
$B\overline{B}$ events~\cite{pipi_babar}. The $CP$-violating parameters are obtained from 
an unbinned extended ML fit to 309540 events. The likelihood function contains 117 parameters 
which are varied in the fit. In order 
to enrich the data sample with the signal events, the event selection requirements 
are lowered in the BaBar analysis and  additional PDF's for background discriminating 
variables (six in all) are included in the likelihood function. The fit
results are ${\cal C_{\pi^+\pi^-}} = -0.21\pm 0.09(stat)\pm 0.02(syst)$ and 
${\cal S_{\pi^+\pi^-}} = -0.60\pm0.11(stat)\pm 0.03(syst)$.

Both Belle and BaBar measurements indicate a large mixing-induces $CP$-violation 
with a significance greater than $5.3\sigma$ and $5.1\sigma$, respectively, for
any values of ${\cal A_{\pi^+\pi^-}}$. Belle also observed large direct $CP$ violation.
The case of no direct $CP$ violation, ${\cal A_{\pi^+\pi^-}} = 0$, is ruled out with 
a significance of $5.5\sigma$. The difference between ${\cal A,S}$ measurements of Belle 
and BaBar, as estimated by Heavy Flavor Averaging Group (HFAG)~\cite{hfag} group, 
constitutes about $2.1\sigma$.

\begin{figure}[h] 
\centering
\includegraphics[width=8.5cm]{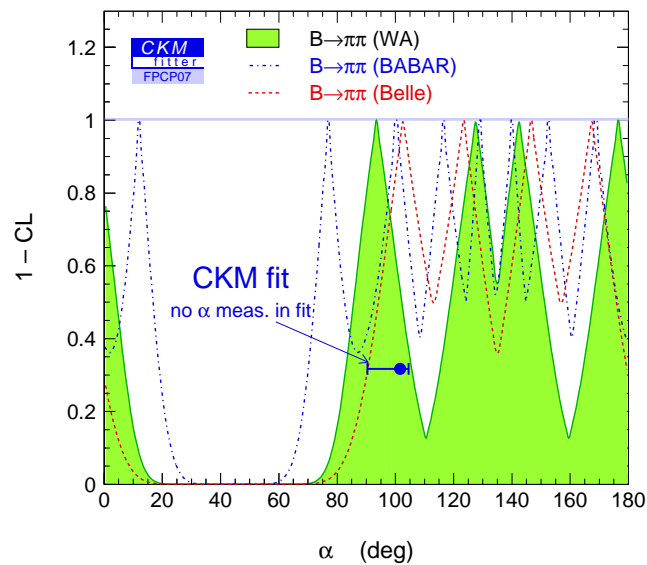}
\caption{1 - C.L.  vs $\phi_2$($\alpha$) obtained from the isospin
analysis of $B\rightarrow \pi\pi$ decays by the CKMfitter group. The dashed curve represents Belle measurements
only; the dot-dashed curve represents BaBar measurements and the hatched region is
a combined constraint from Belle and BaBar.} \label{pipi_constraint}
\end{figure}

The angle $\phi_2$ can be extracted using an isospin relations~\cite{pipi_isospin}. 
The  $SU(2)$ isospin symmetry allows one to relate the amplitudes 
$A^{+-}$, $A^{+0}$, and $A^{00}$ of the 
$B\rightarrow \pi^+\pi^-,\pi^+\pi^0,\pi^0\pi^0$  decays and corresponding 
amplitudes $\bar{A}$ for the charge-conjugated processes as follows
\begin{eqnarray}
\frac{1}{\sqrt 2}A^{+-} + A^{00} = A^{+0} \nonumber \\ 
\frac{1}{\sqrt 2}\bar{A}^{+-} + \bar{A}^{00} = \bar{A}^{-0}. 
\label{isospin}
\end{eqnarray}
These relations lead to two triangles in the complex plane as shown in Fig~\ref{isospin_triangle}.
The angle between these triangles is the phase difference $2\delta\phi_2$.
There are six unknown parameters (five amplitudes, and the angle $\phi_2$) 
and six observables: the branching fractions for $B\ra\pi^+\pi^-$, $\pi^+\pi^0$, and
$\pi^0\pi^0$, the $CP$ parameters ${\cal A}_{\pi^+\pi^-}$ and ${\cal S}_{\pi^+\pi^-}$
and the time integrated asymmetry${\cal A}_{\pi^0\pi^0}$ for $B^0\ra\pi^0\pi^0$ decays.
The angle $\phi_2$ can be determined with an eight-fold ambiguity corresponding to four possible 
orientations of the isospin triangles and the two fold ambiguity from solving Eq.~\ref{eq:phi2eff}.
The confidence level (C.L.) for $\phi_2$  obtained by the CKMfitter group~\cite{ckm_fitter} 
is presented in Fig.~\ref{pipi_constraint}. The curves on this plot correspond to the Belle (BaBar) 
measurements only and to the combined Belle and BaBar constraint. The combined constraint is obtained 
using averaged measurements of the branching fractions and asymmetries.
The eight peaks in the 1-C.L. distribution for BaBar (dot-dashed curve) correspond 
to the eight-fold ambiguity. However the Belle constraint contains only 4 peaks.
This can be explained by the large value of the measured asymmetry 
${\cal A_{\pi^+\pi^-}} = 0$; one of the isospin triangle becomes flat, an apex
of the triangle lies on the base. This leads to merging of the two solutions and to the 
four-fold ambiguity on $\phi_2$. The combined Belle and BaBar constraint on $\phi_2$ 
consistent with the standard model is $93.5^{+12.1}_{-10.0}$ at $68\%$ C.L.

\section{$B^0 \rightarrow \rho^+\rho^-$ \label{sec:rhorho}}
As it was discussed in previous section, $B\rightarrow \pi\pi$ decays 
constrain $\phi_2$ with a four-fold ambiguity. In order
to choose a single solution and further constrain the angle $\phi_2$ one
can exploit other decay channels. In particular, $\phi_2$ can be determined
by measuring $CP$-violating parameters in  $B^0 \rightarrow \rho^+\rho^-$ decays. 
$B^0 \rightarrow \rho^+\rho^-$ channel have several advantages comparing with
$B^0 \rightarrow \pi^+\pi^-$:
\begin{itemize}
\item The branching fraction for $B^0 \rightarrow \rho^+\rho^-$ decays is about 
4.4 times larger than that for $B^0 \rightarrow \pi^+\pi^-$.
\item The size of the penguin amplitude in $B^0 \rightarrow \rho^+\rho^-$ decays
is constrained to be small with respect to the leading tree 
diagram by the small branching fraction of $B^0\ra\rho^0\rho^0$~\cite{babar_rho0rho0}.
\end{itemize}
However there are certain complications in the measurement of $\phi_2$:
\begin{itemize}
\item The relatively large width of $\rho$ mesons ($\sim 150 {\rm MeV}$) leads to 
a substantial combinatorial background in  $\rho^+\rho^-$ decays. 
\item $\rho^+\rho^-$ is a vector-vector final state. The $CP$-violating
parameters receive contributions from a longitudinally polarized state ($CP$-even)
and two transversely  polarized states (an admixture of $CP$-even and $CP$-odd states).
However, recent measurements of the polarization fraction by Belle~\cite{belle_rhorho} and
BaBar~\cite{babar_alpha} show that the longitudinal polarization fraction is near unity
($f_L=0.968\pm 0.023$~\cite{hfag}).
\item The isospin analysis in $B \rightarrow \rho\rho$ decays can be complicated
by a possible contribution of the isospin I = 1 amplitude which might appear due
to the finite width of $\rho$ mesons. Similar to $B^0 \rightarrow \pi^+\pi^-$ decays, 
electroweak penguin amplitudes can contribute to the $B^0 \rightarrow \rho^+\rho^-$ 
decays. However both these contributions are expected to be small~\cite{falk}.
\end{itemize}

Belle measured $CP$-violating parameters using 535 million $B\overline{B}$ pairs.
Similar to $B^0\rightarrow \pi^+\pi^-$ analysis, the measurements are done in two
steps: we first obtain the yields of signal an background components using an 
unbinned extended ML fit to the three-dimensional $(\mmbc,\Delta E, {\cal R})$ 
distribution.
A fit to 176843 events yields $N_{\rho\rho+\rho\pi\pi} = 576\pm 53$. We subsequently 
perform a fit to the $\Delta t$ distribution to determine the $CP$ parameters 
\arhorho\ and \srhorho. The time dependent decay rate for $B \rightarrow \rho^+\rho^-$  
decays is given by Eq. 1. The likelihood function includes the following 
event categories: signal and $\rho\pi\pi$ non-resonant decays, signal decays that have 
at least one $\pi$ meson incorrectly identified (referred to as SCF events), continuum 
background ($q\bar q$), $b\rightarrow c$ background, and charmless  ($b\to u$)  background.
The fit results are ${\cal A}\!=\!0.16\,\pm 0.21 \pm 0.08$ and ${\cal S}\!=\!0.19\,\pm 
0.30 \pm 0.08$.

BaBar analysis~\cite{babar_alpha} is based on a data sample of 347 million $B\overline{B}$ 
pairs.
The signal yield, $f_L$, and $CP$-violating parameters are obtained simultaneously
from an unbinned extended ML fit to 33902 events. The  background discriminating
variables are  $m_{ES}$, $\Delta E$, $\Delta t$, $m_{\pi^\pm\pi^0}$, 
${\rm cos}\theta_\pm$, and $\mathcal N$. PDFs for these variables are 
included in the likelihood function while the event selection requirements
are lowered. The fit results are $N_{\rho\rho} = 615\pm 57$, 
$f_L = 0.977\pm 0.024^{+0.015}_{-0.013}$, ${\cal C}\!=\!-0.07\,\pm 0.15 \pm 0.06$ and ${\cal 
S}\!=\!-0.19\,\pm 0.21^{+0.05}_{-0.07}$. The $CP$-violating parameters measured by
Belle are consistent with those obtained by BaBar. The values of ${\cal A}_{\rho^+\rho^-}$ 
and ${\cal S}_{\rho^+\rho^-}$ are also consistent with no $CP$ violation (${\cal A} = 
{\cal S} = 0$).

\begin{figure}[h]
\centering
\includegraphics[width=8.5cm]{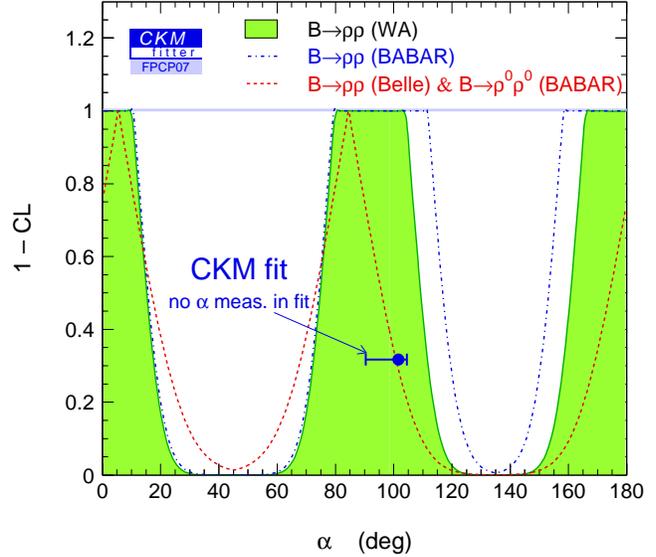}
\caption{1 - C.L.  vs $\phi_2$($\alpha$) obtained from the isospin
analysis of $B\rightarrow \rho\rho$ decays by the CKMfitter group.
The dashed curve represents Belle measurements only; the dot-dashed curve
represents BaBar measurements and the hatched region is
a combined constraint from Belle and BaBar.} \label{rhorho_constraint}
\end{figure}

The angle $\phi_2$ can be constrained using an isospin analysis similar to that
used in $B\rightarrow \pi\pi$ decays. The six underlying parameters are: five 
decay amplitudes for $B\rightarrow \rho\rho$ and the angle $\phi_2$. The observables 
are the branching fractions for $B\ra\rho^+\rho^-$, $\rho^+\rho^0$\,\cite{hfag}, and 
$\rho^0\rho^0$\,\cite{babar_rho0rho0}; the $CP$ parameters ${\cal A}$ and ${\cal S}$ ;
 and the parameter ${\cal A}_{\rho^0\rho^0}$ for $B\ra\rho^0\rho^0$ decays. 
The last parameter is not yet measured, but nevertheless one can constrain $\phi_2$.
We apply the isospin relations to the decay amplitudes corresponding to the longitudinal 
polarization (CP-even state). The branching fractions are multiplied by the corresponding 
longitudinal polarization fractions. We neglect possible contributions from electroweak 
penguins and  $I\!=\!1$ amplitudes~\cite{falk} and possible interference between signal and 
non-resonant components. 
The resulting function $1\!-\!{\rm C.L.}$ obtained by the CKMfitter group is shown 
in Fig.~\ref{rhorho_constraint}. The curves on this plot correspond to the Belle (BaBar)
measurements only and to a combined constraint. The combined constraint is obtained
using averaged measurements of the branching fractions and asymmetries. The distribution 
has more than one peak due to ambiguities that arise when solving for $\phi^{}_2$. 
The ``flat-top''  regions in Fig.~\ref{rhorho_constraint} arise because ${\cal A}_{\rho^0\rho^0}$  
is not measured. The absence of the flat-top regions in Belle measurements can be explained
by the fact that the isospin triangle is not closed; the relatively large measured branching 
fraction of $B^\pm\rightarrow \rho^\pm\rho^0$ decay leads to the squashed triangle.
This results in only a two-fold ambiguity on $\phi_2$. The solution consistent with the standard 
model is $72.5^\circ\!<\!\phi^{}_2\!<\!111.5^\circ$ at 68\%~C.L. Recently, an alternative 
model-dependent approach to extract $\phi_2$ using flavor $SU$(3) symmetry has been 
proposed~\cite{alpha_su3}. This method could potentially give more stringent constraints 
on $\phi_2$.

\section{$B^0 \rightarrow \rho^\pm\pi^\mp$ \label{sec:rhopi}}
An alternative way to measure the angle $\phi_2$ is to perform 
a time-dependent Dalitz plot analysis in $B^0\rightarrow \pi^+\pi^-\pi^0$
decays. It was pointed out by Snyder and Quinn~\cite{snyder_quinn} that the 
Dalitz analysis allows one to determine $\phi_2$ without discrete ambiguities. 
The time-dependent rate for $B^0\rightarrow \pi^+\pi^-\pi^0$ decays
is given by
\begin{eqnarray}
\mathcal{P}(\Delta t, ds_+, ds_-) & & \sim   e^{- |\Delta t| / \tau_{B^0}} \{  
    \left(|A_{3\pi}|^2 + |\overline{A}_{3\pi}|^2\right)  \nonumber \\
   & &  -Q\times [ \left(|A_{3\pi}|^2 - |\overline{A}_{3\pi}|^2\right)
    \cos (\Delta m \Delta t) \nonumber \\
   & & - 2 \mathrm{Im}\left[
                      \frac{q}{p}
                      A_{3\pi}^*
                      \overline{A}_{3\pi}
                     \right]
     \sin (\Delta m \Delta t)]
     \}
    \;,
 \label{eq:rhopi_rate}
\end{eqnarray}
where $Q = +1(-1)$ corresponds to $B^0 (\overline{B}{}^0)$ tags, 
the parameters $q$ and $p$ are the mass eigenstates of neutral $B$ 
mesons with the mass difference $\Delta m$, and the $\tau_{B0}$ is the
$B^0$ meson average lifetime. The amplitudes $A_{3\pi}$($\overline{A_{3\pi}}$)
of the $B^0$($\overline{B}^0$) decays depend on the Dalitz plot
variables $s_{+,-}={(p_{+,-}+p_0)}^2$ and $s_{0}={(p_++p_-)}^2$, where 
$p_{+,-,0}$ are the four-momenta of the $\pi^{+,-,0}$. The amplitudes can be 
factorized as follows
\begin{eqnarray}
 A_{3\pi}(s_+, s_-) & = &
  \sum_{\kappa = (+, -, 0)} 
  T_{J=1}^\kappa F^\kappa (s_\kappa) A^\kappa \;,  \nonumber \\
 \frac{q}{p}
  \overline{A}_{3\pi}(s_+, s_-)
  & = &
  \sum_{\kappa = (+, -, 0)} 
  T_{J=1}^\kappa F^\kappa (s_\kappa) \overline{A}^\kappa \;,
  \label{eq:amplitude}
\end{eqnarray}
where $T_{J=1}^\kappa$, $F^\kappa(s_{\kappa})$, and 
$A^\kappa$($\overline{A}^\kappa$) are helicity distributions, 
lineshapes, and complex amplitudes corresponding to 
$B^0(\overline{B}^0)\rightarrow \rho^+\pi^-, \rho^-\pi^+,\rho^0\pi^0$ 
decays for $\kappa=+,-,0$. The lineshapes $F^\kappa(s)$ is modeled 
as a sum of the $\rho(770)$ resonance and its radial excitations 
$\rho(1450)$, and $\rho(1700)$:
\begin{eqnarray}
F^\kappa(s) = B_{\rho(770)} + \beta B_{\rho(1450)} + \gamma B_{\rho(1700)},
\label{eq:lineshape}
\end{eqnarray}
where $B$ are Breit-Wigner functions and $\beta$ and $\gamma$ are 
the relative complex amplitudes of the two resonances.
Inserting Eq.~(\ref{eq:amplitude}) and Eq.~(\ref{eq:lineshape}) to
Eq.~(\ref{eq:rhopi_rate}) one obtains 26 free parameters 
which are to be determined from a fit to the data.
 
The Belle analysis~\cite{rhopi_belle} is based on a data sample of 449 million
$B\overline{B}$ pairs. We first obtain the signal yield 
from an unbinned extended ML fit to the $M_{\rm bc}-\Delta E$ 
and Dalitz plot distribution. The fit yields $971\pm42$ 
$B^0\rightarrow \pi^+\pi^-\pi^0$ events. The lineshape 
parameters $\beta$ and $\gamma$ are determined from
a time-integrated Dalitz plot fit with a larger
Dalitz plot acceptance. The 26 coefficients are 
subsequently determined from a Dalitz-$\Delta t$ fit 
to the 2824 events in a small signal region. 
The measured parameters also allow to extract a ratio 
$\mathcal{B}(B^0\rightarrow \rho^0\pi^0)/\mathcal{B}(B^0\rightarrow 
\rho^+\pi^-)$ which is measured to be $0.133\pm0.022\pm0.023$.
This value is consistent with a Belle previous measurement
of the  branching fraction of $B^0\rightarrow \rho^0\pi^0$ 
decays using a quasi-two-body approach~\cite{rhopi_quasi}, 
$\mathcal{B}{(B^0\rightarrow 
\rho^0\pi^0)}^{\rm Belle}/\mathcal{B}{(B^0\rightarrow
\rho^+\pi^-)}^{\rm WA}  = 0.130^{+0.049}_{-0.046}$. 

BaBar performed analysis using 375 million $B\overline{B}$ 
pairs~\cite{rhopi_babar}. The 26 coefficients and event yields 
are obtained simultaneously from an unbinned extended ML fit. 
The likelihood function contains in total 68 parameters which are 
varied in the fit. The fit yields $N_{3\pi} = 2067 \pm 68$ candidates.

\begin{figure}[h]
\centering
\includegraphics[width=8.5cm]{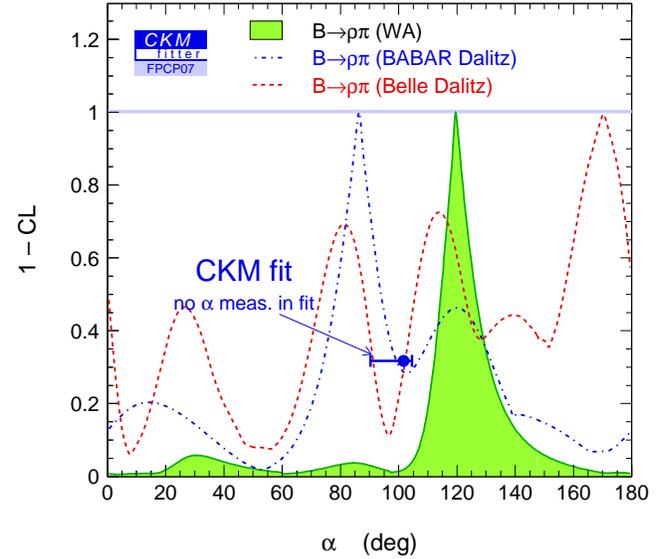}
\caption{1 - C.L.  vs $\phi_2$($\alpha$) obtained from the Dalitz
analysis of $B\rightarrow \rho\pi$ decays by the CKMfitter group.
The dashed curve represents Belle measurements only; the dot-dashed curve
represents BaBar measurements and the hatched region is
a combined constraint from Belle and BaBar.} \label{rhopi_constraint}
\end{figure}

The measured amplitudes $A^\kappa$ and 
$\overline{A}^\kappa$ can be related to the angle $\phi_2$ 
using an isospin relation~\cite{rhopi_isospin} for neutral $B$ decays only
\begin{eqnarray}
e^{2i\phi_2} = \frac{\overline{A}^+ + \overline{A}^- +
2\overline{A}^0}{A^+ + A^- + 2A^0}.
\end{eqnarray}
This relation allows one to determine $\phi_2$ without discrete
ambiguities in the limit of high statistics.
The confidence level for $\phi_2$  obtained by the CKMfitter group is presented in
Fig.~\ref{rhopi_constraint}. The curves on this plot correspond to the Belle 
(BaBar) measurements only and to a global constraint on $\phi_2$ obtained by 
combining Belle and BaBar measurements. The preferred region for the combined constraint 
is around $120^\circ$. 

\begin{figure}[h]
\centering
\includegraphics[width=8.5cm]{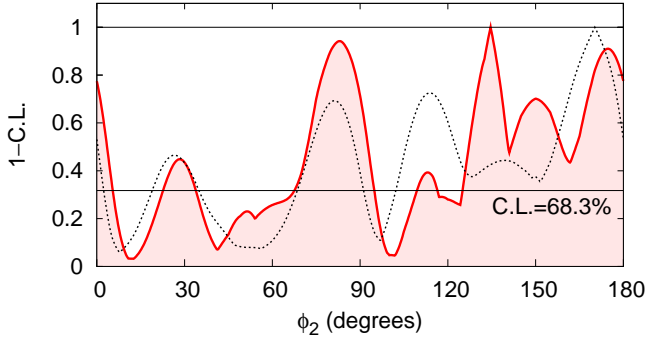}
\caption{1 - C.L.  vs $\phi_2$($\alpha$) obtained by Belle.
The dotted curve represents the Dalitz analysis and the 
solid curve represents the combined Dalitz and an isospin pentagon
analysis.} 
\label{rhopi_belle}
\end{figure}

It is worth mentioning that $\phi_2$ can be further constrained
by including branching fractions $\mathcal{B}(B^0\rightarrow
\rho^\pm\pi^\mp)$, $\mathcal{B}(B^+\rightarrow
\rho^+\pi^0)$, $\mathcal{B}(B^+\rightarrow
\rho^0\pi^+)$ and asymmetries for $\mathcal{A}(B^+\rightarrow
\rho^+ \pi^0)$ and $\mathcal{A}(B^+\rightarrow
\rho^0 \pi^+)$ decays~\cite{hfag} into the isospin analysis (a pentagon isospin analysis). 
As an example, 1 - C.L. curves on  $\phi_2$ obtained by Belle from the Dalitz analysis and 
from the full Dalitz and pentagon combined
analysis are presented in Fig.~\ref{rhopi_belle}. Combined Belle and BaBar 
Dalitz and pentagon constrain is being recently prepared by the CKMfitter group.

\section{$B^0 \rightarrow a_1^\pm\pi^\mp$ \label{sec:a1pi}}

Another channel which allows to measure $\phi_2$ is 
$B^0 \rightarrow a_1^\pm \pi^\mp$. As the final state $a_1^\pm\pi^\mp$ 
is not a $CP$ eigenstate, one has to consider four decay modes with different 
charge and flavor combinations: 
$B^0 \rightarrow a_1^\pm \pi^\mp$ and $\overline{B}{}^0 \rightarrow a_1^\pm \pi^\mp$.
The decay rates can be written as~\cite{a1pi}

\begin{eqnarray}
\mathcal{P}_{a_1^\pm\pi^\mp}(\Delta t)  & = & (1\pm \mathcal{A}^{a_1 \pi}_{CP}) 
\frac{e^{-|\Delta t|/\tau_{B^0}}}{4\tau_{B^0}}\{1 - Q\times \nonumber \\ 
       &  & [(C_{a_1\pi}\pm\Delta C_{a_1\pi}) \nonumber \cos(\Delta m_d \Delta t) -  \nonumber \\
       &  & (S_{a_1\pi}\pm\Delta S_{a_1\pi})\sin(\Delta m_d \Delta t)]\}, 
\label{eq:rate}
\end{eqnarray}
where $\Delta t = t_{\rho\pi} - t_{tag}$ is the proper-time difference
between the fully reconstructed and the associated $B$ decay, and $Q = +1(-1)$ 
corresponds to $B^0 (\overline{B}{}^0)$ tags. The parameters 
$S_{a_1\pi}$ and $C_{a_1\pi}$ are associated with mixing-induced $CP$ 
violation (related to $\phi_2$) and flavor-dependent direct $CP$ violation, 
respectively. The parameters $\Delta S_{a_1\pi}$ and $\Delta C_{a_1\pi}$ are 
$CP$-conserving. $\Delta C_{a_1\pi}$ describes the asymmetry between the rates 
$\Gamma (B^0 \to a_1^+\pi^-) + \Gamma (\overline{B}{}^0 \to a_1^-\pi^+)$
and $\Gamma (B^0 \to a_1^-\pi^+) + \Gamma (\overline{B}{}^0 \to a_1^+\pi^-)$.
$\Delta S_{a_1\pi}$ depends in addition on difference in strong phases 
between the amplitudes contributing to $B \to a_1\pi$ decays.

The parameters $C$, $\Delta C$, $S$, and $\Delta S$ have recently been measured 
by BaBar collaboration using a data sample of 384 million $B\overline{B}$ 
pairs~\cite{a1pi_babar}.
The likelihood function includes the following components: signal, generic
$B\overline{B}$ background, continuum $q\bar{q}$ background, 
$B^0\rightarrow a_2^\pm(1320)\pi^\mp$, and non-resonant $\rho\pi\pi$. 
The fitting observables are $m_{\rm ES}$, $\Delta E$, a Fisher discriminant $F$ for
continuum suppression, $m_{a_1}$, $\Delta t$ and an  angle between the flight 
direction of the bachelor pion from $B$ meson and normal to the plane of the $3\pi$
 resonance calculated in the $3\pi$ rest frame. An unbinned extended ML fit to
29300 events yields $608\pm 53$ signal events and the following parameters:
$C_{a_1\pi} = -0.10\pm0.15\pm0.09$, $\Delta C_{a_1\pi} = 0.26\pm0.15\pm0.07$, 
$S_{a_1\pi} = 0.37\pm0.21\pm0.07$, $\Delta S_{a_1\pi} = -0.14\pm0.21\pm0.06$, and
${\mathcal A}_{CP}^{a_1\pi} = -0.07\pm0.07\pm0.02$. These measurement indicate no
direct and mixing-induced $CP$ violation in $B^0\rightarrow a_1^\pm \pi^\mp$ decays.
As shown in ~\cite{a1pi}, the effective angle $\phi_2^{\rm eff}$ can be calculated 
as follows 
\begin{eqnarray}
\phi_2^{\rm eff} = \frac{1}{4}[\arcsin(\frac{S_{a_1\pi} + \Delta S_{a_1\pi}}
{\sqrt{1 - (S_{a_1\pi} + \Delta S_{a_1\pi})^2}} + \nonumber \\
\arcsin(\frac{S_{a_1\pi} - \Delta S_{a_1\pi}}{\sqrt{1 - (S_{a_1\pi} - \Delta S_{a_1\pi})^2}}].
\label{api_rate}
\end{eqnarray}
$\phi_2^{\rm eff} = \phi_2 + \delta\phi_2$ is measured to be $78.6^\circ \pm 7.3^\circ$. 
The extraction of $\phi_2$ can be performed using an SU(3) flavor symmetry~\cite{a1pi}. 
However it can not be done at the moment as branching fractions for SU(3)-related decays 
have not yet been measured.

\section{Summary}
\begin{figure}[h]
\centering
\includegraphics[width=8.5cm]{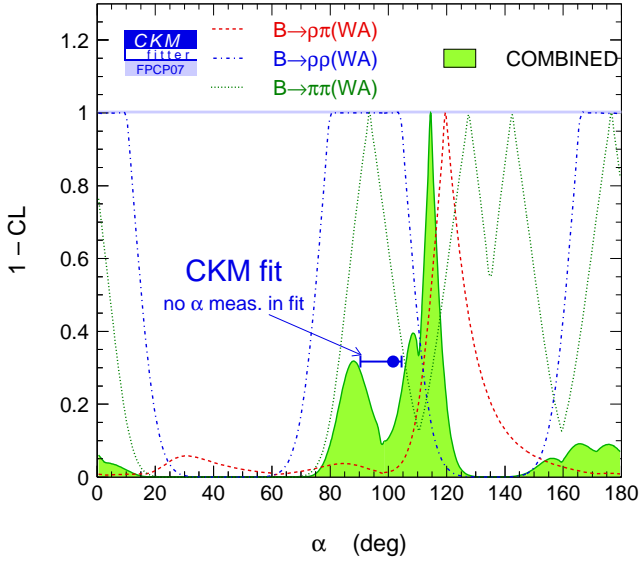}
\caption{World average constraints on $\phi_2$($\alpha$) obtained from
an isospin analyses of $B\rightarrow \pi\pi$  (dotted curve) and
 $B\rightarrow \rho\rho$ (dot-dashed curve) decays, and from Dalitz analysis
of $B\rightarrow \rho\pi$ decays (dashed curve). The hatched region is
a combined constraint for these three channels.} \label{wa_constraint}
\end{figure}

We have discussed the measurements of the CKM phase angle $\phi_2$ using 
$B\rightarrow \pi\pi,\,\rho\rho,\,\rho\pi$, and $a_1^\pm\pi^\mp$ decays.
The combined constraint for the first three channels is presented in 
Fig.~\ref{wa_constraint}. There are two 
preferred regions around $88^\circ$ and $115^\circ$. This can be explained
by the fact that the $\phi_2$ preferred region obtained from $B\rightarrow \rho\pi$ Dalitz
analysis doesn't agree well with that obtained from the isospin analysis
of $B\rightarrow \pi\pi$ and $B\rightarrow \rho\rho$ decays. The angle $\phi_2$ can be 
constrained as $(114.5^{+4.4}_{-8.3})^\circ$
at 68\%~C.L. (solution around the main peak) and  $80.0^\circ\!<\!\phi^{}_2\!<\!122.7^\circ$
at 90\%~C.L. As it can be seen, there is no 'stringent' constrain at 90\%~C.L. with the 
current data. More data are awaited to improve our knowledge on $\phi_2$.

{}


\begin{thebibliography}{}

\bibitem{ckm}
M.\ Kobayashi and T.\ Maskawa, Prog.\ Theor. Phys. {\bf 49}, 652 (1973);
N.\ Cabibbo, Phys.\ Rev.\ Lett.\ {\bf 10}, 531 (1963).

\bibitem{angle_definition}
 Angles $\phi_1$, $\phi_2$, and $\phi_3$ in Belle are referred to as $\alpha$, 
 $\beta$, and $\gamma$  in BaBar.

\bibitem{chargeconjugate}
Charge-conjugate modes are included throughout this paper
unless noted otherwise.

\bibitem{snyder_quinn}
A.\ E.\ Snyder and H.\ R.\ Quinn, Phys.\ Rev.\ D {\bf 48}, 2139 (1993).


\bibitem{fox_wolfram}
  G.~C.~Fox and S.~Wolfram,
  Phys.\ Rev.\ Lett.\  {\bf 41}, 1581 (1978).


\bibitem{KSFW} S.\,H.\,Lee {\it et al.}, Phys.\ Rev.\ Lett.~{\bf 91}, 261801 (2003).


\bibitem{asym_def}
 BaBar is using a definition ${\cal C} = -{\cal A}$.


\bibitem{pipi_belle}
  H.~Ishino {\it et al.}  (Belle Collaboration),  Phys.\ Rev.\ Lett.\  
  {\bf 98}, 211801 (2007).

\bibitem{pipi_babar}
  B.~Aubert {\it et al.}  (BABAR Collaboration),  Phys.\ Rev.\ Lett.\  {\bf 99}, 021603 (2007).

\bibitem{hfag}
Heavy Flavor Averaging Group, August 2006,
\url{http://www.slac.stanford.edu/xorg/hfag/}.

\bibitem{pipi_isospin}
  M.~Gronau and D.~London,  Phys.\ Rev.\ Lett.\  {\bf 65}, 3381 (1990).

\bibitem{ckm_fitter} J.\ Charles {\it et al.} (CKMfitter Group), Eur.\ Phys.\ J.\ C {\bf 41}, 
1 (2005).

\bibitem{babar_rho0rho0}
B.\,Aubert {\it et al.} (BABAR Collaboration), 
Phys.\ Rev.\ Lett.\  {\bf 98}, 111801 (2007).

\bibitem{belle_rhorho}
A.~Somov {\it et al.} (Belle Collaboration) ,
Phys.\ Rev.\ Lett.\  {\bf 96}, 171801 (2006).


\bibitem{babar_alpha} 
B.\,Aubert {\it et al.} (BABAR Collaboration),
Phys.\ Rev.\ Lett.\ {\bf 95}, 041805 (2005);
Phys.\ Rev.\ Lett.\ {\bf 93}, 231801 (2004); 
hep-ex/0607098.

\bibitem{falk} 
A.\ Falk {\it et al.}, 
Phys.\ Rev.~D {\bf 69}, 011502(R) (2004).

\bibitem{alpha_su3}
  M.~Beneke, M.~Gronau, J.~Rohrer and M.~Spranger,
  Phys.\ Lett.\ B {\bf 638}, 68 (2006).


\bibitem{rhopi_belle}
  A.~Kusaka {\it et al.}  (Belle Collaboration),
  Phys.\ Rev.\ Lett.\  {\bf 98}, 221602 (2007).

\bibitem{rhopi_quasi}
  J.\ Dragic  {\it et al.} (Belle Collaboration), Phys.\ Rev.\ D {\bf 73}, 111105 (2006).

\bibitem{rhopi_babar}
  B.~Aubert {\it et al.}  (BABAR Collaboration),
  Phys.\ Rev.\  D {\bf 76}, 012004 (2007).

\bibitem{rhopi_isospin}
 H.\ J.\ Lipkin, Y.\ Nir, H.\ R.\ Quinn, and A.\ E.\ Snyder,
  Phys.\ Rev.\  D {\bf 44}, 1454 (1991); M.\ Gronau, 
  Phys.\ Lett.\ B {\rm 265}, 389 (1991).

\bibitem{a1pi} M.~Gronau and J.~Zupan,
  Phys.\ Rev.\  D {\bf 73}, 057502 (2006)

\bibitem{a1pi_babar}
  B.~Aubert {\it et al.}  (BABAR Collaboration),
  Phys.\ Rev.\ Lett.\  {\bf 98}, 181803 (2007).

\end{thebibliography}
\end{document}